\documentclass[10pt,preprint2]{aastex}

\usepackage{booktabs}
\usepackage{lscape}
\usepackage{graphicx,longtable}
\usepackage{array}
\usepackage{amsmath}
\usepackage{amssymb}
\usepackage{subfigure}

\begin{document}

\title{Impact of new data for neutron-rich heavy nuclei  on theoretical models for $r$-process nucleosynthesis}

\author{Toshitaka Kajino}
\affil{ Division of Theoretical Astronomy, NAOJ, 2-21-1 Osawa, Mitaka, Tokyo, 181-8588, Japan}
\affil{ Department of Astronomy, The University of Tokyo, 7-3-1 Hongo, Bunkyo-ku, Tokyo, 113-033, Japan}
\affil{School of Physics and Nuclear Energy Engineering, Beihang University, Beijing 100191, China}

\author{Grant J. Mathews}
\affil{ Center for Astrophysics, University of Notre Dame, Notre Dame, IN 46556, USA}
\affil{ Division of Theoretical Astronomy, NAOJ, 2-21-1 Osawa, Mitaka, Tokyo, 181-8588, Japan}

\begin{abstract}
Current models for the $r$ process are summarized with an emphasis on the key constraints from both nuclear physics measurements and astronomical observations. 
In particular, we analyze the importance of nuclear 
 physics input such as beta-decay rates; nuclear
masses; neutron-capture cross sections; beta-delayed neutron
emission; probability of spontaneous fission, beta- and neutron-induced fission,
fission fragment mass distributions; neutrino-induced reaction cross sections, etc.
We highlight the effects on models for $r$-process nucleosynthesis of newly measured $\beta$-decay half-lives, masses, and spectroscopy  of neutron-rich nuclei
near the $r$-process path.
We overview $r$-process nucleosynthesis in the neutrino driven wind above the proto-neutron star in core collapse supernovae
along with the possibility of magneto-hydrodynamic jets from rotating supernova explosion models.  We also 
consider the possibility of neutron star mergers as an $r$-process environment.
A key outcome of newly measured nuclear properties far from stability is the degree of shell quenching for neutron rich isotopes near the
 closed neutron  shells.  This leads to  important constraints on the sites for $r$-process nucleosynthesis in which  freezeout occurs
on a rapid timescale.
~\\
\\
\vskip .4 in
\end{abstract}
\vskip .2 in
\date{\today}

\maketitle


\vskip .8 in
\section{Introduction}

Rapid neutron-capture ($r$-process) nucleosynthesis
is responsible for the origin of approximately
half of the elements heavier than iron and
is the only means to produce the naturally occurring
radioactive heavy actinide elements such as Th and U.
However, in spite of more than a half century of study and observational progress
[e.g.~\cite{Sneden08}], 
the astrophysical sites for $r$-process nucleosynthesis
have not yet been unambiguously identified
(for reviews see, e.g., \cite{Arnould07, Martinez08,Thielemann11}).
Although many candidate sites have been proposed \citep{Mathews90}, at present only neutron star mergers (NSMs) or core collapse supernovae (CCSNe)
appear to be well suited as an $r$-process site.  Nevertheless,
there is still  no consensus as to the correct astrophysical site.  Indeed, it is undoubtedly the case that more than
one astrophysical site has contributed to the observed Solar-System $r$-process abundances \citep{Wehmeyer15,Shibagaki16}.

At least part of the reason for the difficulty in identifying the $r$-process site has been the lack of experimental data for the relevant 
neutron rich nuclei.  However, in recent years the first systematic direct measurements  of  properties of the very neutron rich nuclei relevant to $r$-process nucleosynthesis 
have become available. 
 In this review we identify in part the impact that new measurements are having on the search for the site for the $r$-process.

Notwithstanding the difficulties in finding a suitable astronomical environment,
the physical conditions for the $r$-process are well constrained \citep{Burbidge57} by simple nuclear physics considerations as described in following  sections.
It is evident that the $r$-process occurs via a sequence of near equilibrium rapid neutron captures
and photo-neutron emission reactions far on the neutron-rich side of stability.
This equilibrium is established with a maximum abundance strongly peaked
on one or two isotopes far from stability.
The relative abundance of $r$-process elements is then determined
by the relative $\beta$-decay rates along this $r$-process path.,
i.e. slower $\beta$-decay lifetimes result in higher abundances.
At least part of the reason for the difficulty in finding the astrophysical site
for the $r$-process stems from the fact that it lies so far
from the region of stable isotopes
where until recently there has been little experimental data on nuclear masses, structure, and $\beta$-decay rates.

In this context, it is of particular interest
that recently many masses and $\beta$-decay half-lives have been measured.  For example, 
up to 110 neutron-rich isotopes of the elements from Rb to Sn have measured \citep{Lorusso15} at the RIKEN Radioactive Isotope Beam Factory.  
These  isotopes encompass the neutron closed shell at $N=82$.  The new half-lives show evidence for  new  systematic features 
and exhibit  a persistence of shell effects. The new data have helped to calibrate both beta-decay rates and new mass tables [e.g. \citep{Nishimura12,Moller12,Lorusso15}].  These measurements have direct implications for $r$-process calculations and reinforce the notion that the second ($A\approx130$) and the rare-earth-element ($A\approx165$) abundance peaks may result from the freeze-out of an 
$(n,\gamma)\leftrightarrows (\gamma,n)$ equilibrium. In such an equilibrium, the new half-lives are important factors determining the relative abundance of rare-earth elements, and allow for a more reliable discussion of the $r$ process environment.

Many of these newly measured isotopes are near or directly on the $r$-process path.
As such, they are of particular interest as they determine
the $\beta$-flow toward the important $r$-process peak at $A=130, N=82$.
Thus, they regulate the ability of models for the $r$-process
to form heavier elements \citep{Otsuki03,Shibagaki16}.
It is of particular interest, therefore,
to examine the impact of these new rates on the specific numerical models for $r$-process that we are most involved with.
For further insight the reader is referred to many other recent reviews and sensitivity studies   [e.g. \cite{Brett12, Surman14, Mumpower16}] in the context of more schematic hot, cold, exponential time-dependence, adiabatic expansion, etc.~$r$-process models.  

We begin in  Section \ref{obsconstraints} with a brief summary of  key constraints from astronomical observations.  In Section \ref{basics}
we review the basic physics of the $r$ process and highlight the  required input nuclear data.
Then, in Section \ref{nucleardata} we summarize the input data available.  In Section \ref{models} we overview the main models for the $r$-process.
In Section \ref{impacts} we summarize the key directions the field has taken as a result of the present new data, and also describe some implications for
understanding the galactic chemical evolution of the $r$ process in Section \ref{GCE}.
In Section \ref{needed} we summarize our (admittedly biased) view of what new data may be most desired in future measurements.
We conclude in Sections  \ref{concl} with a summary of the  status of $r$-process models based upon the current data.

\section{Observational Constraints}
\label{obsconstraints}
The present understanding of the origin of $r$-process elements has been helped greatly by the detailed elemental abundance distributions observed  in $r$-process enhanced metal-poor
stars in the Galactic halo \citep{Sneden08}.  A key point is that the elemental abundances for these stars appear to identically match the Solar-System $r$-process abundances. This apparent ``universality'' in $r$-process abundances argues \citep{Mathews90,Argast00,Argast02,Argast04} in favor of a single $r$-process environment that occurred early in the history of the Galaxy.  

However, it is anticipated that universality may not extend to the elements Sn, Sb, I, and Cs, making the detection of these elements in metal-poor stars of the utmost importance to determine the exact conditions of individual $r$-process events. 
Moreover, the scatter in the [Eu/Fe] abundances at low metallicity argues that the $r$-process is a rare event \citep{Mathews90,Mathews92,Ishimaru99, Argast02,Argast04}.  Indeed, the two most popular models for the $r$-process, NSMs or magneto-hydrodynamic jets (MHDJ), are indeed rare events compared to the event rate of normal core-collapse supernovae.  More recently, it has been noted \citep{Roederer14} that stars that exhibit  $r$-process enhanced abundances do not exhibit the same enhancement in  $\alpha$ elements (although the $\alpha$ elements are indeed enhanced).  This suggests that the $\alpha$ elements are not enhanced to the same degree as the  $r$ process when the $r$-process occurs  in CCSN or NSMs.  This supports the idea that the $r$-process is a rare event.

Recently, new insight has been gained from the observation of $r$-process elements in dwarf galaxies \citep{Okamoto08,Frebel10,Frebel12, Ji16, Roederer16}.  For the most part, dwarf galaxies show $r$-process abundances similar to that of the Galactic halo \citep{Frebel12}.  However, the recent observations \citep{Ji16,Roederer16} of seven stars in the dwarf galaxy Reticulum II,  show evidence of a rare single event that ejected a large mass ($\sim 0.2$ M$_\odot$) of 
$r$-process elements.  Such an event is strongly suggestive of a neutron star merger that is  inherently a rare event and  capable of ejecting a large mass of $r$-process elements.  However, some magneto-hydrodynamically driven jet models may also produce a comparable mass.  

In these  metal-poor stars in the Reticulum II, enhanced $\alpha$ elements Mg and Ca are also detected.   Very interestingly these are  at the same level as Sr, Y, Zr and Ba, i.e. [X/Fe] $\approx 0.5$ to $1$, except for extremely enhanced [Eu/Fe] $\approx 1.5$ to  $2$  \citep{Roederer16}.  It was previously noted \citep{Roederer14} that metal-poor stars in the Milky Way halo that exhibit $r$-process enhanced abundances do not necessarily exhibit the same enhancement in lighter elements.  Although Na and Al are indeed not enhanced, the $\alpha$ elements like Mg, Si and Ca are enhanced.  The existence of $\alpha$ elements in the metal-poor stars of  Reticulum II and the Milky Way halo suggests that the $\alpha$ elements were ejected to the same degree as the $r$ process elements when the $r$-process occurred in the rare event like a NSM or MHDJ in the early Galaxy.  Note, however, that it is difficult for NSMs  to eject lighter elements  with $A < 100$ as will be discussed in Section 6.  

The lack of light elements in NSMs  could also affect the dust formation \citep{Takami14}.   Dust grains have difficulty forming  in the ejecta from NSMs due to the low number density of the lighter  elements.   Lighter elements like carbon and silicon, for example,  are required for the condensation of silicon carbide X-grains that are classified to be supernova grains.  Such grains may not be able to form in the ejecta of NSMs.

Another significant recent bit of observational evidence concerns the `kilonova' light curve from short duration gamma-ray bursts \citep{Tanvir13, Berger13}.  A faint 'kilonova` transient  following the burst  is attributed to the decay of neutron-rich radioactive species generated during the merger of two neutron stars.  The near infrared emission also 
exhibits excess flux possibly due to the high opacity of the newly synthesized heavy elements.  This provides evidence of active heavy-element nucleosynthesis in NSMs supporting this environment as a site for $r$-process nucleosynthesis.

\section{Basics of the  $r$-Process}
\label{basics}

The $r$ process involves  a sequence of rapid neutron captures in an explosive environment \citep{Burbidge57,Mathews85}.   Although many sites have been proposed for the $r$-process, whatever the environment, it can be shown that  the Solar-System $r$-process abundances are well reproduced by beta-decay flow in a system that is in approximate $(n,\gamma) \leftrightarrows (\gamma,n)$ equilibrium.
Hence, the relative abundances of isotopes of a given element are determined by  nuclear statistical equilibrium (NSE) as described by the nuclear Saha equation \citep{Saha21}.
\begin{eqnarray}
\frac{n(Z,A) }{n(Z,A+1)} &=&  \frac{1}{n_n}  \biggl(\frac{2 \pi \mu kT}{ h^2}\biggr)^{3/2} 
\nonumber \\
&&\times \frac{G_{A}G_{n}}{G_{A+1} } e^{-Q_n/kT}~~,
\label{rproc}
\end{eqnarray}
where $\mu$ is the reduced mass of the neutron plus isotope $^{A}Z$, $h$ is Planck's constant, $k$ is the Boltzmann constant, and $T$ is the temperature.  The quantity $G_A$ is the partition function for nucleus, $^{A}Z$,  $Q_n$ is the neutron capture $Q$ value for isotope $^{A}Z$ (or equivalently the neutron separation energy for the nucleus $^{A+1}Z$),  and $n(Z,A)$ represents the number density of an isotope $^{A}Z$.   Note, however, that this formula neglects a small correction \citep{Mathews11} for the difference between Maxwellian and Planckian distribution functions for the photons.

Equation (\ref{rproc}) defines a sharp peak in abundances for one (or a few) isotopes within an isotopic chain.   The flow of beta decays along these peak isotopes is then known as the $r$-process path.  

The location of the $r$-process path peak is roughly identified \citep{Burbidge57} by the condition that neutron capture ceases to be efficient once  $n(Z,A+1)/n(ZA)^<_\sim 1$.  
Taking the logarithm of Eq.~(\ref{rproc}) and inserting the numerical terms, the $r$-process path can be identified by the following relation
\begin{eqnarray}
\biggl(\frac{Q_n}{kT}\biggr)_{\mbox{\footnotesize path}} &=& 
  2.30 \biggl(35.68 + \frac{3}{2}\log{(\frac{kT}{\mbox{\footnotesize MeV}})} \nonumber \\
  &- & \log{(\frac{n_n}{\mbox{\footnotesize cm}^{-3}})} \biggr)~~.\nonumber
 \label{path} 
 \end{eqnarray}
The  elemental abundances $n(Z,A)$  along this path are then determined by the flow of beta decays,
\begin{equation}
\frac{dn(Z,A)}{dt} = \lambda_{Z-1} n(Z,A-1) - \lambda_Z n(Z,A)~~,
\end{equation}
where the total beta decay rate of each element along the path is given by the weighted sum of  beta decay rates for each isotope $\lambda_\beta(Z,A) = 1/\tau_\beta(Z,A)$:
\begin{equation}
\lambda_Z = \sum_A n(Z,A) \lambda_\beta(Z,A)
\label{betaflow}
\end{equation}
For a typical $r$-process temperature of $T_9\sim 1 $, the requirement that the $r$-process path reproduce the observed abundance peaks at $A = 80, ~130$ and $195$, implies that the $r$-process path halts at waiting points in the beta flow near the neutron closed-shell nuclei $^{80}$Zn, $^{130}$Cd and  $^{195}$Tm.  For a neutron density sufficiently high ($n_n \ ^>_\sim 10^{20}$ cm$^{-3}$) so that the neutron capture rates exceed the beta-decay rates for these isotopes, the peak abundances along the $r$-process path must be for isotopes with $Q_n \sim 1-3$ MeV, and thus $(Q_n/kT)_{path} \sim 10-30$.  
 
 This constraint on $Q_n$, however, concerns the conditions near "freezeout" when the final neutrons are exhausted at the end of the $r$-process.   At this point, the system falls out of NSE and nuclei along the $r$-process path decay back to the line of stable isotopes.  

 Earlier in the $r$ process the neutron densities can be quite high and the $r$-process path shifted to more  neutron-rich nuclei.    For example, in the neutrino driven wind (NDW) models of \cite{Woosley94}, the $r$-process conditions begin with a neutron density  of $n_n \approx 10^{27}$  cm$^{-3}$ and a temperature of $T_9 \sim 2$.  The density is also much higher ($> 10^{32}$ cm$^{-3}$) when the  material is first ejected from the proto-neutron star.   Such conditions can also be achieved for an $r$ process which occurs during NSMs \citep{Freib99,Rosswog99,Rosswog00,Korobkin12,Piran13,Rosswog13}.  
 
 Of course, as the $r$ process freezes out, one must make a detailed accounting of the full $r$-process reaction network, i.e.
 \begin{eqnarray}
 \frac{dn(Z,A)}{dt} &=& n(Z,A-1)\phi_n \sigma_{n,\gamma}(Z,A-1) \nonumber \\ &+&  n(Z,A+1) \phi_\gamma \sigma_{\gamma,n}(Z,A+1) \nonumber \\
 &+&  n(Z-1,A) \lambda_{\beta}(Z-1,A) \nonumber \\
 &+& {\rm ~terms~with~} (n,p), (n,\alpha), (p,\gamma), (\alpha,\gamma), \nonumber \\
 &+& (n,{\rm fission}), (\beta,n), (\beta, {\rm fission}),~~{\rm etc.} \nonumber \\
 &-& n(Z,A)[\phi_n \sigma_{n,\gamma}(Z,A) + \lambda_\beta(Z,A) \nonumber \\
&+&\phi_\gamma \sigma_{\gamma,n}(Z,A) + \cdot \cdot \cdot ]
 \end{eqnarray}
 where $ \phi_n$ and $\phi_\gamma$ are the time-dependent neutron and photon fluxes, respectively.  Recent work \citep{Mumpower12, Mumpower15a, Mumpower15b, Mumpower16} has demonstrated  that nuclear properties of a few isotopes in the range of $Z \approx 53-60, N \approx 100-115$ can have a dramatic effect on the final freezeout abundances for the rare-earth peak. 
 
\section{Nuclear Input Data}
\label{nucleardata}

\subsection{Nuclear Reaction Network}

The nuclear reaction network is a key part  of the nucleosynthesis simulations.  
An $r$-process network typically  consists of more than $4000$ isotopes, including neutrons, protons,
and heavy isotopes with atomic number $Z \leq 100$
[for example, see Table 1 in \cite{Nishimura06}].  As noted above, nuclear reactions are most important as the system falls out of $(n,\gamma)$ equilibrium where residual neutron captures can smooth the odd-even effect in the abundance distribution and shift the final abundances.  They are also important in the build up of light nuclei to form the $r$-process seed nuclei \citep{Sasaqu05a}

\subsection{Nuclear Reaction Rates}
The most important nuclear reaction rate for the $r$-process is the neutron capture rate for isotopes along the $r$-process path when the system falls out of $(\gamma,n)$ equilibrium.
In addition to neutron capture, one should also consider other possible reactions related to the $r$-process
involving two- and three-body reactions or decay channels.  Also, 
one should include electron capture as well as positron capture
and screening effects for all of the relevant charged particle reactions.

\subsection{New Measurements of Neutron Capture Rates}

Available experimentally determined and theoretical neutron-capture reaction rates for $r$-process nucleosynthesis
are maintained in REACLIB \citep{Cyburt10}, and the Karlsruhe Database of Nucleosynthesis in Stars (KADoNiS) \citep{Dillmann06, Rauscher12}.   As theoretical estimates become better constrained by measurements near the $r$-process path, a better identification of the $r$-process site will follow.  

Unfortunately, neutron capture rates along the $r$-process path are exceedingly difficult to measure.  There is, however, the possibility \citep{Reifarth14} that the  combination of a radioactive beam facility, an ion storage ring and
a high flux reactor would allow a direct measurement of inverse neutron capture  reactions for isotopes with half lives down to $\sim$minutes. The idea is that radioactive ions  pass through a neutron
target.  A storage ring of radioactive ions could be used 
to enhance the luminosity.

Even without direct measurements,  however, useful information can be  inferred  \citep{Chiba08b,Jones11,Kozub12} using $(d,p)$ reactions near the $r$-process path at $A=130, N=82$.  For example, in \cite{Kozub12}, direct-semi-direct $(n,\gamma)$  cross section calculations were made, based for the first time on experimental data. The uncertainties in these cross sections were  thus reduced by orders of magnitude compared to that of  previous estimates.  

Another possibility is to  infer  $(\gamma,n) $ cross sections using virtual photons from Coulomb excitation with a radioactive ion beam.  This technique has been successfully demonstrated in\citep{Utsunomiya12}.   In that paper a  $\gamma$-ray strength function method was devised to determine radiative neutron capture cross sections for unstable nuclei along the valley of beta-stability. This method is based on the $\gamma$-ray strength function which interconnects radiative neutron capture and photoneutron emission within the statistical model. The method was applied to several unstable nuclei such as $^{93,95}$Zr, $^{107}$Pd, and $^{121,123}$Sn. This method offers a versatile application extendable to unstable nuclei far from the stability when combined with Coulomb dissociation experiments at RIKEN-RIBF and GSI.

\subsection{Theoretical Neutron Capture Rates}
The nuclear reaction flow in the $r$ process occurs in the vicinity of the neutron drip line. 
There are two  main theoretical approaches for neutron capture reactions.  These are:  1) via a compound nucleus (including resonances) as in the Hauser Feshbach estimates; 2)  direct capture and  semi-direct (DSD) processes. For most applications of $r$-process nucleosynthesis,  nuclear cross sections have been based upon  a simple estimate of the direct and semi-direct cross sections in terms of pre-equilibrium $\gamma$  emission \citep{Akkermans85}, or in the context of  Hauser-Feshbach  theory, [e.g., \cite{Koning04, Young92, Rauscher00, Cyburt10}]. Such an approach can be justified when it is applied to  nuclei in the vicinity of the stability line. However, in the neutron-rich region relevant to the $r$ process, the neutron separation energies are diminished, so the compound nuclei may not have enough level density to compete with the compound elastic process. In this case, the compound capture cross section may be suppressed, and direct capture becomes dominant even at low energies.  

Normally, the direct process is not  very important because its cross section is much smaller than the compound capture cross sections.  However, in \cite{Mathews83}  it was noted that far from stability  where the level density is low, the direct capture process could be the dominant  mode of  neutron capture reaction for  the $r$-process.

In \cite{Chiba08} the DSD components of the neutron capture cross sections were calculated for a number  of tin isotopes by employing a single-particle potential (SPP) that gives a good reproduction of the known single-particle energies (SPEs) over a wide mass region. The results were compared with the Hauser-Feshbach  contribution in the energy region of astrophysical interest. Their calculations showed that the Hauser-Feshbach component drops off rapidly for the isotope $^{132}$Sn and toward more neutron-rich nuclei, whereas the DSD component decreases gradually and eventually becomes the dominant reaction mechanism.  In \citep{Chiba08} the reason for the difference in the isotopic dependence between the Hauser-Feshbach and DSD components was discussed, and its implication for $r$-process nucleosynthesis was given.

This result is consistent with those of previous studies, but the dependence of the DSD cross section on the target mass number is a feature of their  SPP that gave a smooth variation of SPEs. As a consequence, the direct portion of the DSD components gave the largest contribution to the total ($n, \gamma$) cross section for neutron-rich isotopes below a few MeV. Therefore, the direct capture process modifies significantly the astrophysical ($n, \gamma$) reaction rates. The semi-direct component, however, gives a negligible contribution to the astrophysical reaction rates, but its impact is significant above several MeV.

 Valuable studies of the impact of varying theoretical neutron capture rates in $r$-process models can be found in \cite{Mumpower16}.  In those papers a  Monte Carlo variation of Hauser-Feshbach neutron capture rates within the context of several mass models was explored.    Crucial isotopes in the vicinity of the $r$-process peaks at $A=130$ and 195 were identified and also in the vicinity of the rare-earth peak whose measurement would be most effective in reducing the uncertainties in $r$-process abundance calculations.

\subsection{Nuclear Masses }
Experimentally determined masses \citep{Audi95,Audi03,Wang12} should be  adopted if available.
Otherwise, the theoretical predictions for nuclear masses are necessary.  At present there are many  available 
theoretical mass estimates far from stability.  A good resource for nuclear masses can be found at http://nuclearmasses.org.

Theoretical mass tables for  $r$-process nuclei are  mainly based upon three approaches.  One is  the macroscopic/microscopic method based upon a liquid droplet formula
plus  shell corrections.  The most popular adaptation of this is the finite range droplet model (FRDM) \citep{Moller95,Moller12}.   Another variant of the macroscopic/microscopic approach is the phenomenological hybrid KTUY model \citep{Koura00,Koura05}.  A third is the DZ model \citep{Duflo99} based upon a parametrization of multipole moments of the nuclear Hamiltonian.  In a sense  the DZ  is more fundamental than the macroscopic/microscopic models.  However,  it is not strictly a microscopic theory, since no explicit nuclear interaction appears in the formulation.   

At the next level would be masses based upon the extended Thomas Fermi random phase approximation (ETSFI)  plus Strutinsky integral semi-classical approximation to a Hartree Fock (HF)  approach \citep{Aboussir95, Pearson96}.  The most microscopic extrapolations generally available of masses for neutron rich nuclei are those based upon the Skyrme Hartree-Fock Bogolyubov (HFB) method.   This is a fully variational, approach with single-particle energies and pairing treated simultaneously and on the same footing.  Some recent formulations include: HFB-19, HFB-21 \citep{Goriely10}; Gogny HFB \citep{Goriely09a}; the Skyrme-HFB \citep{Goriely09b, Chamel08};  HFB-15 \citep{Goriely08}; HFB-14 \citep{Goriely07}.  

A good comparison of the relative merits of each approach can be found in 
\cite{Pearson06}.  All approaches give a reasonable fit to known nuclear masses.  However, there can be large deviations as one extends the mass tables to unknown neutron rich nuclei.  Hence, there is a need for experimental mass determinations for neutron rich nuclei.  

A recent study has been made \citep{Martin16} of the impact of nuclear mass uncertainties based upon six Skyrme energy density functionals based on different optimization protocols.  Uncertainty bands related to mass modeling for $r$-process abundances were determined for  realistic astrophysical scenarios.  This work highlights the critical role  
of experimental nuclear mass determinations for understanding the site for $r$-process nucleosynthesis.

\subsection{New Experimental Masses}
There are now active programs at CERN, GSI, RIKEN, JYFL, ANL, and NSCL to measure experimental masses on and near the $r$-process path.  In particular, masses adjacent to the classical waiting-point nuclide $^{130}$Cd have been measured \citep{Atanasov15} using the Penning-trap spectrometer ISOLTRAP at ISOLDE/CERN.   That  work reported   a significant deviation ($\sim 400$ keV) from earlier mass estimates  based upon  nuclear beta-decay endpoint data. The new measurements indicated  a  reduction of the $N=82$ shell gap below the doubly magic nucleus $^{132}$Sn.   

A similar conclusion was reached in \cite{Hakala12} based upon mass measurements at JYFL.   This has a significant impact on models for the $r$ process in either CCSNe or 
NSMs as reported in that paper.  

\subsection{Nuclear Structure Studies}
The level structure of nuclei along the $r$-process path is important both as a means to determine the partition functions and as a means to test the strength of shell closures.  Recently a number of studies have been completed \citep{Watanabe13, Simpson14, Taprogge14} in the neighborhood of the $N=82, A=130$ $r$-process peak.  In particular,  the first ever studies \citep{Watanabe13} of the  level structure of the waiting-point nucleus $^{128}$Pd   and $^{126}$Pd have been completed.  That study  indicated that the shell closure at the neutron number $N=$82 is fairly robust.  Hence, there is conflicting  evidence between the nuclear masses and nuclear spectroscopy as to the degree of shell quenching near the $N=82$ closed shell.  It will be  important to clarify this point as it has important implications for the site of $r$-process nucleosynthesis as discussed below.

\subsection{Beta-Decay Rates}
The  $\beta$-decay rates, particularly at  waiting point nuclei, constitute  one of the most important nuclear physics inputs to  nucleosynthesis calculations in  the $r$-process. 
Theoretical investigations of the beta decay of isotones with neutron magic number of $N = 82$ have been done by various methods including the shell model \citep{Zhi13}, quasiparticle random-phase approximation (QRPA)/finite-range droplet model (FRDM) \citep{Moller03}, QRPA/extended Thomas-Fermi plus Strutinsky integral (ETFSI) \citep{Borzov00}], and Hartree-Fock-Bogoliubov (HFB) + QRPA \citep{Engel99} calculations as well as in the continuum quasiparticle random-phase approximation (CQRPA) \citep{Borzov03}. The half-lives of nuclei obtained by these calculations are rather consistent with one another, and especially in shell-model calculations experimental half-lives at proton numbers $Z = 47, 48,$ and $49$ are well reproduced \citep{Martinez99}.

For the $\beta$ decays at N = 126 isotones, however, half-lives obtained by various calculations differ from one another \citep{Langanke03,Grawe07}. First-forbidden (FF) transitions become important for these nuclei in addition to the Gamow-Teller (GT) transitions in 
contrast to the case of $N = 82$. 

A strong suppression of the half-lives has been predicted in \cite{Borzov03} for  $N = 126$ isotones due to the FF transitions. Most shell-model calculations of the $\beta$-decay rates of $N = 126$ isotones have been done with only  the contributions from the GT transitions included \citep{Langanke03, Martinez01}. Moreover, as noted below experimental data for the $\beta$ decays in this region of nuclei are not yet available. The region near the waiting point nuclei at $N = 126$ is therefore called the Òblank spotÓ region.

In \cite{Suzuki12} , $\beta$ decays of $N = 126$ isotones were studied by taking into account both the GT and FF transitions to evaluate their half-lives. 
Shell-model calculations were done with the use of shell-model interactions based upon modified G-matrix elements that reproduce well the observed energy levels 
of the isotones with a few (two to five) proton holes outside $^{208}$Pb \citep{Steer08,Rydstrom90}. 

In \cite{Marketin15} the impact of first-forbidden transitions on decay rates was 
studied using a fully self-consistent covariant density functional theory (CDFT) framework to provide a table of $\beta$-decay half-lives and $\beta$-delayed neutron emission probabilities, including first-forbidden transitions.   This works demonstrated that there is a significant contribution of the first-forbidden transitions to the total decay rate of nuclei far from the valley of stability.  This also brings better agreement with experimentally determined half lives as discussed below.

In addition to ground state $\beta$ decay, at the high temperatures of the $r$-process environment decay can proceed through thermally excited states.  In \cite{Famiano08} a calculation was made  to evaluate the possible effects of the $\beta$-decay of nuclei in excited-states on the astrophysical $r$-process. Single-particle levels were calculated in the FRDM model with quantum numbers determined based upon their proximity to Nilsson model levels. The resulting rates were  used in an $r$-process network calculation. Even though  the decay rate model was simplistic,  this work did provide a measure  of the possible effects of excited-state $\beta$-decays on $r$-process freeze-out abundances.  The main result of that work was that in the more massive nuclei, the speed up of the decay rates  in the approach to closed shells tended to exaggerate the underproduction of nuclei below the nuclear closed shells as discussed below.

\subsection{New Experimental Beta Decay Rates}
There are currently many active programs to measure $\beta$-decay rates for nuclei near the $r$-process path \citep{Dillmann03, Pfeiffer01,Hosmer05, Montes06,Hosmer10, Quinn12, Madurga12,Benlliure12, Benzoni12, Domingo13, Kurtukian14, Morales14, Lorusso15}.  Even so, not all  half-lives of nuclei most relevant to the $r$-process  have been measured in the vicinity of either the $N = 50$ or  $N = 82$ closed shells. Moreover, though experiments  have provided valuable information for half-lives approaching the first and second $r$-process peaks, so far no experimental half-lives are available for $r$-process nuclei at the important rare-earth peak or the  third r-process peak at the $N = 126$ closed shell.  However, next-generation facilities such as FRIB, ARIEL, RIBF, RISP
SPIRAL2, ISOLDE upgrade, RIBLL, and FAIR will hopefully soon extend the list of measured isotopes to heavier nuclei on and  near the $r$-process path.

\subsection{Beta-Delayed Neutron Emission}
Beta-delayed neutrion emission is particularly important for the freezeout of the $r$-process.   Recently,  beta-delayed neutron emission probabilities of neutron rich Hg and Tl nuclei have been measured  \citep{Caballero16} together with beta-decay half-lives for 20 isotopes of Au, Hg, Tl, Pb, and Bi in the region of neutron number $N \ge 126$. These are the heaviest nuclear species for which  neutron emission has been observed. 

Although not directly on the $r$-process path, these measurements have provided information with which to evaluate the viability of nuclear microscopic and phenomenological models for  the high-energy part of the beta-decay strength distribution. Indeed, this study indicated that there is no global beta-decay model that provides satisfactory beta-decay half-lives and
neutron branchings on both sides of the $N = 126$ shell
closure. There was, however,  a slight preference for the Hartree-Bogoliubov model  of \cite{Marketin15}.

\subsection{Fission Barriers and Fission Fragment Distribution}
In $r$-process models with a very high neutron-to-seed ratio (such in the ejecta from neutron star mergers) the $r$-process path can proceed until neutron-induced or beta-induced fission terminates the beta flow at $A \sim 300$.  Determining where this occurs can significantly impact the yields from $r$-process models \citep{Eichler15, Shibagaki16}.  Unfortunately there are no measurements of fission barriers or fission fragment distributions (FFDs) for nuclei heavier than $^{258}$Fm \citep{Schmidt12}.

This is a major uncertainty in all calculations of fission recycling in the $r$-process.  \cite{Shibagaki16} considered a FFD model based upon the KTUY model plus a two-center shell model to predict both symmetric and asymmetric FFDs with up to three components.  As such,  fissile nuclei could span a wide mass range (A=100-180) of fission fragments as demonstrated below.

On the other hand, the $r$-process models of \cite{Korobkin12} were  mostly based upon a simple two fragment distribution as in \cite{Panov01} (or alternatively the prescription of \cite{Kodama75}).  The assumption of only two fission daughter nuclei  tends to place a large yield near  the second $r$-process peak leading to a distribution that looks rather more like the solar $r$-process abundances.  In contrast, the FFDs of \cite{Goriely13} are based upon a rather sophisticated SPY revision \citep{SPY} of the  Wilkinson fission model \citep{Wilkins76}.  The main ingredient of this model is that the individual potential of each fission fragment is obtained as a function of its axial deformation from tabulated values. Then a Fermi gas state density is used  to determine the main fission distribution.   This leads to a  FFDs with  up to four humps.   

An even more important  aspect is the termination of the $r$-process path and the number of fissioning nuclei that contribute to   fission recycling and the  freezeout of the $r$-process abundances.  
The $r$-process path in \cite{Shibagaki16} proceeded rather below the fissile region until nuclei with $A \sim 320$, whereas the $r$-process path in \citep{Goriely13} terminates at $A \approx 278$ [or for a maximum $\langle Z \rangle$ for \cite{Korobkin12}]. Moreover, \cite{Shibagaki16} found that only $\sim10$\% of the final yield comes from the termination of the $r$-process path at N = 212 and Z = 111,  while almost 90\% of the $A=160$ came  from the fission of  more than 200  different parent nuclei mostly via beta-delayed fission. 
On the other hand, the yields of \cite{Goriely13} that are almost entirely due to a few  $A \approx 278$ fissioning nuclei with a characteristic four hump FFD.  As noted below, this has a dramatic impact on the final   $r$-process abundance distribution.

\section{Current Models for the $r$-Process}
\label{models}

In spite of its simplicity, as noted above, the unambiguous identification of the sites for  $r$-process nucleosynthesis has remained elusive.  
The required high neutron densities and short explosive  time-scales ($\sim$ seconds) suggest that both  CCSNe and NSMs
 are viable candidates for $r$-process nucleosynthesis.   Observations \citep{Sneden08} showing the appearance of heavy-element $r$-process abundances early in the history of the Galaxy seem to favor the short stellar lifetime of CCSNe as the $r$-process site.  However, identifying the $r$-process site in models of CCSNe has been difficult \citep{Arnould07, Thielemann11}.  

\subsection{Neutrino Driven Wind Model}
For a number of years,  a favored model for  $r$-process nucleosynthesis was  in the neutrino driven wind (NDW)  above the newly forming neutron star in CCSNe \citep{Woosley94}.  In this model, as a neutron star is formed by the collapse of the iron core of a massive star, the cooling of the proto-neutron star is characterized by the release of $\sim 10^{53}$ ergs in neutrinos on a timescale of $\sim 10$ sec.  The interaction of these neutrinos with material behind the outgoing supernova shock generates a hot bubble that helps to drive the explosion \citep{Bethe95}.  It also leads to the ablation of material from the proto-neutron star into the hot bubble.  This  has been dubbed a neutrino-driven wind.  

Although the original formulation was quite successful, in subsequent calculations the NDW  has been shown \citep{Fischer10,Hudepohl10} to be inadequate as an $r$-process site when modern neutrino transport methods are employed along with a stiff nuclear equation of state \citep{Warren16} as required by observations \citep{Demorest10, Antoniadis13} of neutron stars with masses as large as $2.01 \pm 0.04$ M$_\odot$.  

Of particular importance in this regards.  It has been shown \citep{Olson16} that by adopting a Skyrme density functional equation of state that is consistent with constraints on the symmetry energy from the combination of isobaric analog states, pygmy resonances, and heavy ion collisions, that the decrease in neutrino energies and luminosity results.  Hence, the likelihood of a NDW $r$-process has been affected by measurements that constrain the nuclear EoS and in particular, the density dependence of the nuclear symmetry energy.
The result is that the  desired conditions of high entropy and  neutron-rich composition  \citep{Otsuki00, Otsuki03} do not occur in the neutrino energized wind. Nevertheless, it is quite likely that the so-called "weak $r$-process" occurs  in the NDW producing neutron rich nuclei up to about $A \sim 125$ \citep{Wanajo13, Shibagaki16}.

\subsection{Neutron Star Mergers}  
Indeed, the difficulties in reproducing the $r$-process abundances have motivated many new studies of NSMs either as  $NS+NS$ or $NS + BH$ binaries [e.g.~\cite{Goriely11, Korobkin12,  Piran13, Rosswog13,Rosswog14,Goriely13,Wanajo14, Nishimura16}].  

The ejected matter from  NSMs is very neutron-rich ($\langle Z/A \rangle \equiv Y_e \sim 0.1$).  This means that   the $r$-process path can proceed along the neutron drip line all the way to the  region of fissile nuclei ($A \approx 300$).  Indeed, in such models, fission recycling occurs.  That is, after the $r$-process terminates by beta-induced or neutron-induced fission, the fission fragments continue to experience neutron captures until  fission again terminates the $r$-process path thereby  repeating the process.  After a few cycles   the abundances become dominated by the fission fragment distributions and not as much by the beta-decay flow near the closed shells.  Hence, nuclear data to constrain theoretical models of nuclear fission modes and mass distributions become very important.  

Recent studies \citep{Goriely11, Korobkin12, Goriely13, Wanajo14, Nishimura16} have indicated  that the $r$ process in NSMs produces a final abundance pattern that can be similar to the solar 
$r$-process abundances, but only for heavier A$>$130 nuclei.  However, it is possible \citep{Shibagaki16} that the main  effects of fission recycling may be to fill in the bypassed abundances in the main $r$-process as discussed below.   

However, the distribution of  nuclear fission products can affect the final abundance pattern significantly. 
One must carefully  extrapolate fission fragment distributions (FFDs) to the vicinity of the $r$-process path [cf.~\cite{Martinez07,Erler12}].  It can be argued \citep{Shibagaki16}, however, that by incorporating the expected broad distribution of fission fragments, the effect of the neutron closed shells becomes smoothed out, thereby providing a means to fill in the isotopes bypassed in the main $r$ process.

The study of \cite{Shibagaki16} made use of  self consistent  $\beta$-decay rates,  $\beta$-delayed neutron emission probabilities, and  $\beta$-delayed fission probabilities  taken from 
 \cite{Chiba08} and based upon fits to known fragment distributions.  The spontaneous fission rates and the $\alpha$-decay rates were  taken from  \cite{Koura04}. For these rates   $\beta$-delayed fission  is the dominant nuclear fission mode near the termination of the $r$ process \citep{Chiba08}. However, this is a phenomenological model.  Current microscopic calculations tend to produce lower fission barriers and narrower fragment distributions as discussed below.

\subsection{Magneto-hydrodynamic Jet Models}
Although the NDW model does not seem to be a good source for the main $r$ process,  one scenario for  $r$-process nucleosynthesis in CCSNe remains viable.  It is the magneto-hydrodynamic jet (MHDJ) supernova model \citep{Nishimura06, Winteler12, Nishimura15}.  In this model magnetic turbulence leads  the ejection of neutron rich material into a jet.  As the jet transports this  neutron-rich  material away from the star it can undergo $r$-process  nucleosynthesis in a way that avoids the problems associated with neutrino interactions in the  NDW model.  Moreover, the required conditions of the $r$-process environment
(timescale, neutron density, temperature, entropy, electron fraction, etc.) are well accommodated in this model. Such jet models also have the advantage that they can naturally provide a site for a strong but rare $r$-process early in the  early history of the Galaxy as required from astronomical observations.

The main features of the MHDJ supernova  model 
are described in detail in  \cite{Nishimura06} and \cite{Winteler12}.  We briefly outline  the content of each model here.
In  \cite{Nishimura06}  two-dimensional MHD simulations were  carried out from the onset of the core collapse to the shock propagation to the silicon-rich layers ($\sim 500$ ms after bounce). Thereafter,  the $r$-process nucleosynthesis was calculated in the later phase by employing the two kinds of time extrapolations of the temperature and density starting  from the composition produced  the explosion.   

A  jet-like explosion could  be formed from 
 the combined effects of rapid rotation (ratio of rotational energy to gravitational energy $T/W \approx 0.5$\%) and a  strong initial magnetic field ($\sim 10^{13}$ G).  This jet  had a lower  electron fraction than in  spherical explosions. As the ejected material with low $Y_e$ in the jet emerged  from the silicon layers, an $r$ process occurred that was  able to reproduce the solar  $r$-process abundance distribution up to  the third ($A \approx 195$) peak. 

\cite{Winteler12} expanded upon earlier MHDJ simulations by utilizing a three-dimensional magneto-hydrodynamic core-collapse supernova model and also including an approximate treatment of the neutrino transport.    As in the two-dimensional calculations, in order to form  the  bipolar jets,  a  rare progenitor configuration characterized by a high rotation rate ($T/W \approx 0.8$\%) and a large magnetic field ($5 \times 10^{12}$ G) was required.  This magnetic field was amplified to $\sim 10^{15}$ G during collapse by the conservation of magnetic flux.   As in \cite{Nishimura06}  the low $Y_e$ material ejected in the jet underwent $r$-process nucleosythesis that reproduced the second and third peaks of the solar $r$-process element distribution.  

However, all of these jet simulations tend to underproduce nuclides just below and above the $r$-process abundance peaks.  This tendency is affected by the new measurements of masses and $\beta$-decay rates near the closed neutron shells along the $r$-process path as discussed below.

\subsection{Collapsar $r$ Process}
There has been some interest \citep{Fujimoto06, Fujimoto07, Fujimoto08, Surman08, Ono12, Nakamura13, Nakamura15}  in the possibility of $r$-process nucleosynthesis  in the relativistic jets associated with the collapsar (failed supernova) model for gamma-ray bursts.  See \cite{Nakamura13,Nakamura15} for a recent review.  Collapsars are  a favored model for the formation   of observed  long-duration gamma-ray bursts (GRBs). In the collapsar model \citep{Woosley93, Paczynski98, MacFadyen99,MacFadyen01,Popham99,Aloy00,Zhang03} the central core of a massive star collapses to a black hole.  Angular momentum in the progenitor star, however, leads to the formation of a heated accretion disk around the nascent black hole.  Magnetic field amplification and heating from the pair annihilation of thermally generated neutrinos  emanating from this accretion disk  can then launch material in a polar funnel region leading  to an outflow of neutron-rich
matter from the accretion disk into a relativistic jet along the polar axis.  

A large volume of work has explored  the formation of such collapsars 
[e.g. \cite{Takiwaki04, Kotake04, Sawai04, Obergaulinger06, Suwa07, Burrows07, Takiwaki09, Taylor11},    See also references in \cite{Kotake06}], 
and the development of the associated relativistic jets [e.g.
\cite{MacFadyen99, MacFadyen01,Popham99,Aloy00,Zhang03, Proga03, Hawley06, Mizuno07, Fujimoto06, McKinney07, Komissarov07, Nagataki07, Barkov08,Nagakura11}].

The work of \cite{Nakamura15} utilized a model from  \citet{Harikae09}  for slowly rotating collapsar models in axisymmetric special relativistic magneto-hydrodynamics (MHD).   A method  \citep{Harikae10}  was also applied  to compute the  detailed neutrino-pair heating by ray-tracing neutrino transport to explore whether  the collapsar model is indeed capable of generating  the high entropy per baryon and neutron-rich material required for an $r$-process  in a high-Lorentz-factor jet heated via neutrino-pair annihilation.   
Hydrodynamic studies  of material in the heated  jet along with  the associated nucleosynthesis were evolved out to the much later times and lower temperatures associated with the $r$ process.  

\begin{figure}[htbp]
	\begin{center}
 \includegraphics[width=0.8\hsize]{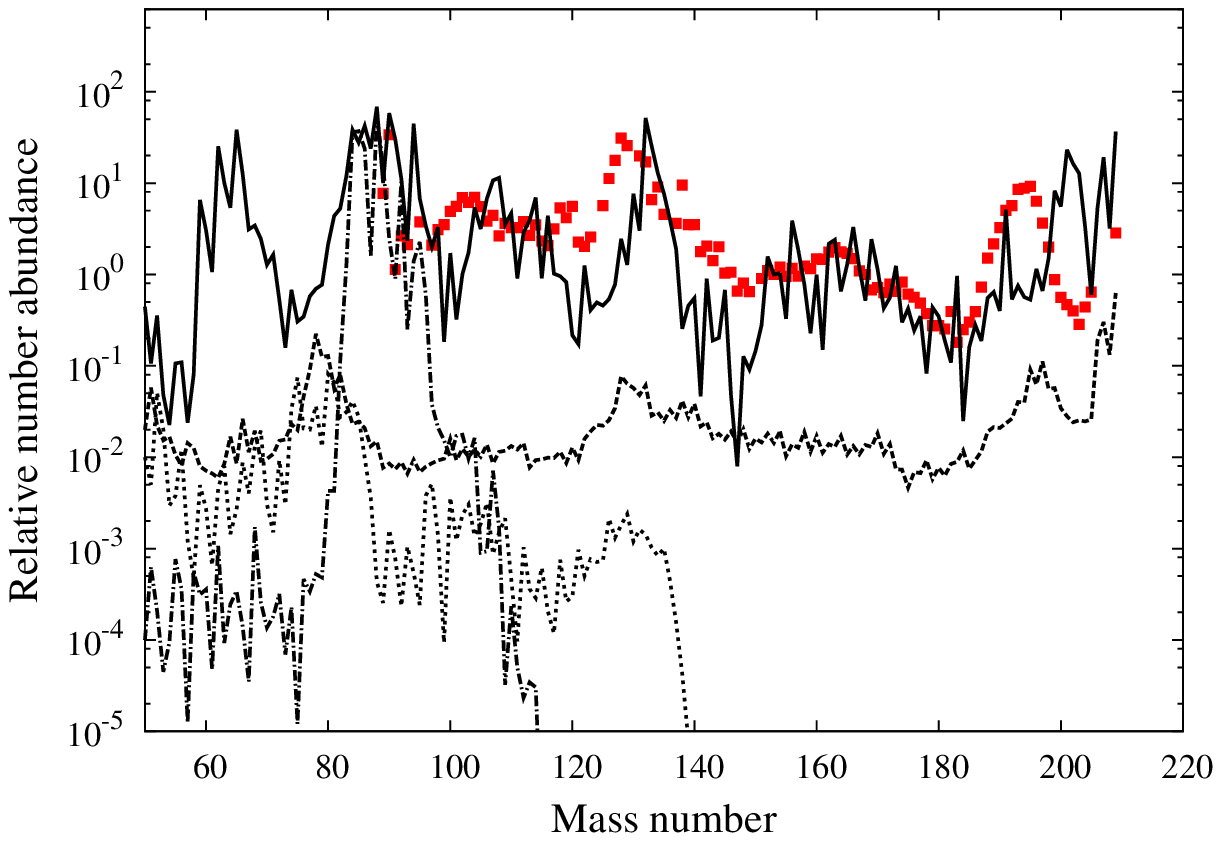}
    \caption{
    Summed mass-weighted  abundance distribution (solid line) for the 1289 ejected tracer particles ejected in the collapsar jet [from \cite{Nakamura15}].  
    These are compared with the Solar-System $r$-process abundances (points)  \citep{Kaeppeler89} and normalized at $^{153}$Eu.
Also shown are arbitrarily normalized  abundances of elements synthesized in individual trajectories with high entropy $S/k = 1000$ (dashed), intermediate entropy $S/k = 100$ (dotted), and low entropy $S/k = 25$ (dash-dotted).}
\label{fig1}
	\end{center}
\end{figure}

It was found \citep{Nakamura15} that  this environment could indeed produce an $r$-process-like abundance distribution.  However, the very rapid time scale and high entropy caused the abundances to differ from the solar abundance distribution as shown in Figure \ref{fig1}.  This is an extreme example of a model with a very rapid freezeout.  As noted below, new information on nuclear masses and $\beta$-decay rates near the $r$-process path now place important constraints on such models.

\subsection{The $tr$-Process}
It has been pointed out \citep{Boyd12} that the abundance patterns observed for the stars that do not fit the standard $r$-process template might  be produced by stars that are sufficiently massive that their core collapse at first produces neutron stars.  Subsequently, however,  the  infall onto the proto-neutron star causes a collapse to a black hole.  This is the so-called fallback supernova. Stars in this class span a mass range from roughly 25 to 40 solar masses \citep{Heger02} for low-metallicity stars. When the neutron star collapses to a black hole the ongoing $r$-process ceases, terminating either when the $r$-processed regions are swallowed by the black hole or when the electron antineutrinos fall below the event horizon \citep{Sasaqu05}. Thus, this truncated $r$-process, or $tr$-process, nucleosynthesis could terminate at different stages what would have been a normal neutrino-driven wind $r$-process, depending on the precise time at which the black hole prevented further $r$-process production or emission of nuclides into the interstellar medium. In this paradigm, therefore,  the delayed collapse to the black hole, combined with  the difficulties in observing the higher mass rare-earth elements, could suggest  a cutoff in the $r$-process distributions observed around $A\approx 165$.
The implementation of this scenario then simply assumes that mass layers  that produce the lighter $r$-process nuclei are ejected prior to those  that produce the heavier
$r$-process nuclides as in NDW models \citep{Woosley94}. However, any setting within a core-collapse supernova that satisfies this condition could lead to a truncated $r$-process. 

 For this process as in the models noted above the nucleosynthesis is sensitive to the nuclear masses, beta decay rates, and neutron separation energies, but in this case the relevant atomic mass numbers are for light nuclei, $A< 100$.  As a possible additional benefit of the $tr$-process, it was noted in \cite{Boyd12}  that in some cases the stars produced no nuclides in the $A =130$ peak or beyond. This could have the effect of enhancing  the yields of the lightest $r$-process nuclides relative to the main $r$ process.  Hence, this could provide an alternative means to fill  in nuclei in the $A=  110-120$ region. Indeed, evidence in metal poor stars of  the production  of nuclides in the $ A= 110-120 $ mass region and lighter may indicate
  a $tr$-process origin.  This conclusion, however, is very dependent upon nuclear properties of the light neutron-rich nuclei along the $r$-process path and requires a much more detailed simulation of the relevant astrophysics.


\section{Impact of New Data on Models for the $r$ Process}
\label{impacts}

\subsection{Impact on the MHD Jet model}

\begin{figure}[htbp]
	\begin{center}
 \includegraphics[width=0.8\hsize]{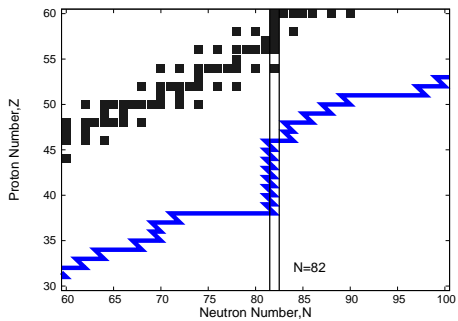}
    \caption{
    Illustration \citep{Shibagaki16} of the $(N,Z)$ path of $r$-process nucleosynthesis (blue line) for nuclei with $A \sim 90$ -- $150$ in the vicinity of the $N=82$ neutron closed shell just before freezeout of the abundances.	 Black squares show the stable isotopes.}
\label{fig2}
	\end{center}
\end{figure}

There is a persistent problem in the MHDJ  model, or any model [e.g.~\cite{Otsuki03}] in which the $r$-process elements are produced on a short time scale via the rapid expansion of material away from the neutron star.  Most such models underproduce isotopic abundances just below and above the $r$-process abundance peaks.  
Figure \ref{fig2} from \cite{Shibagaki16} illustrates why this occurs.  

This figure  shows an example of a typical calculated $r$-process path near the $N=82$ neutron closed shell just before freezeout when the  neutrons are exhausted and the synthesized nuclides  begin to beta decay back to the region of stable isotopes.   Neutron captures and photo-neutron emission proceed in equilibrium for nuclei with a neutron binding energy of about 1-2 MeV.  For $r$-process models with a rapid transport time, the density diminishes rapidly so that a sudden freezeout occurs close to the $r$-process path.  However, above and below a closed neutron shell the $r$-process path shifts abruptly toward the closed shell from below (or  away from the closed shell for higher nuclear masses).  This shifting of the $r$-process path causes isotopes with $N = 70-80$ ($A \sim$110-120) or $N=$90-100 ($A \sim$140-150) to be bypassed in the beta-decay flow.   Indeed, this was a consistent feature in the original realistic NDW models of \cite{Woosley94}.  This effect was apparent in many $r$-process calculation since the 1970s [cf.~review in \cite{Mathews85}].

Although it has been speculated for some time [e.g.~\cite{Woosley94,Pfeiffer01,Farouqi10}] that this could be due to quenching of the strength of the shell closure or beta-decay rates near the closed neutron shell, this explanation seems unlikely.    Recent measurements \citep{Nishimura11,Lorusso15} 
of beta-decay half lives near the $r$-process path have confirmed \citep{Nishimura12} that the reason for this discrepancy cannot be attributed to uncertainties in  nuclear beta-decay properties of nuclei along the $r$-process path.  

\begin{figure*}[htbp]
	\begin{center}
		\includegraphics[width=\hsize]{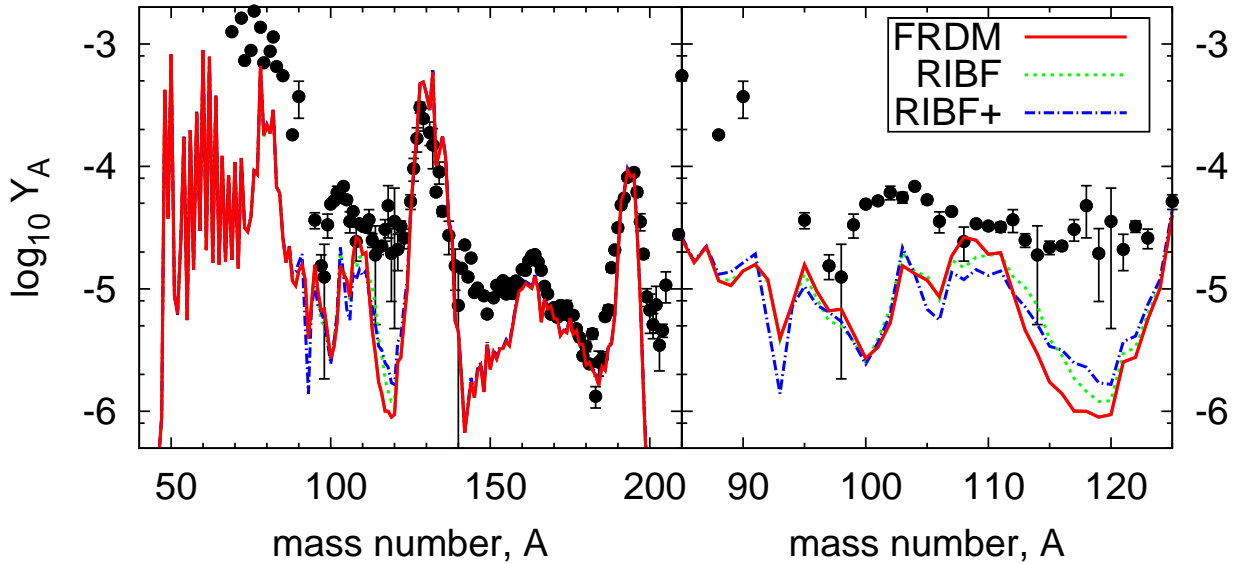}
		\caption{
		Integrated mass averaged total final abundance distributions
		of $r$-process elements from \cite{Nishimura12} based upon the  MHDJ supernova model
		 \citep{Nishimura06}.
		Red solid, green dotted, and blue dashed lines correspond
		to results from using the FRDM (standard), RIBF, and RIBF+ rates, respectively.
		Abundances of Solar-System $r$-process isotopes \citep{Arlandini99}
		are represented by black dots with error bars.} 
		\label{tot-abund}
	\end{center}
\end{figure*}

In \cite{Nishimura12} the $r$-process abundances were calculated in an  MHD jet-like explosion based upon the  two-dimensional magneto-hydrodynamic simulation of  \cite{Nishimura06}.
This study  explored the impact of new beta-decay rates on models with rapid transport.  
The ejecta were evolved with 23 tracer particles to describe the evolution
of the thermodynamic state variables.
These were then post-processed to obtain the nucleosynthesis yields.

The $r$-process calculations  \citep{Nishimura12} were based upon
three different nuclear reaction networks. 
One of the networks utilized only the FRDM theoretical rates \citep{Moller95}
from the REACLIB compilation \citep{Rauscher00}.
The other two (RIBF and RIBF+) utilize the new experimental $\beta$-decay half-lives
of 38 neutron-rich isotopes from \rm{Kr} to \rm{Tc}
and two versions of the theoretical FRDM $\beta$-decay rates for the other isotopes.
The  RIBF network replaced the FRDM rates with the new measured rates where possible.
The third network (RIBF$+$) was based on the RIBF network and FRDM rates
with modified $Q$-values for $(n,\gamma)$ and reverse reactions given by:
\begin{equation}
	Q_{\rm{+}} =
	\begin{cases}
	Q - 0.3 ~\rm{[MeV]} &(~~97 \le A \le 103) \\
	Q + 0.5 ~\rm{[MeV]} &(104 \le A \le 107) \\
	Q + 1.0 ~\rm{[MeV]} &(108 \le A \le 115)
	\end{cases}
\label{qvalue}
\end{equation}
where $Q$ is the theoretical Q-value (in MeV) obtained from the FRDM \citep{Moller95}.
These  modified $Q$-values were adopted because they led to a better fit
to the measured $\beta$-decay lifetimes away from stability.

In particular, the underproduction of isotopes near $A = 120$
became slightly less pronounced relative to predictions based upon
the FRDM rates when the new measured rates were employed. 

Figure \ref{tot-abund} from \cite{Nishimura12} shows the final integrated abundance distributions
from all of the trajectories.
These are compared to the Solar-System $r$-process abundance distribution
of \cite{Arlandini99}.
Here the effect of new rates becomes apparent.
Although it was hoped that the newly measured $\beta$-decay rates
in this mass region might shift the $\beta$-flow equilibrium
thereby filling in the low abundances near $A \sim 120$.
Figure \ref{tot-abund} shows that the abundances
in the $A = 110$ -- $120$ region are only slightly enhanced.
Thus, although the new rates provide a little assistance
in enhancing the abundances below the abundance peak,
they did not alleviate this problem. This suggested that a further modification of the $r$-process paradigm is required.

\begin{figure}[htbp]
	\begin{center}
 \includegraphics[width=0.8\hsize]{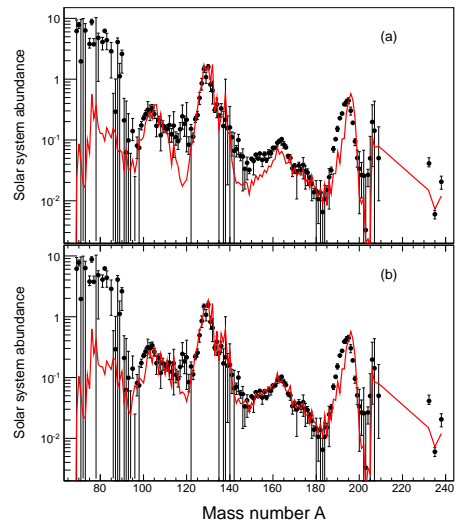}
    \caption{
    Comparison between schematic high-entropy wind model calculations \citep{Lorusso15} without the new beta decay rates (upper plot) and a calculation with the new rates (lower plot).}
\label{figL4}
	\end{center}
\end{figure}

For example, \cite{Lorusso15} were able to avoid 
the underproduction in a schematic high-entropy  slower  outflow model that summed  over several  wind trajectories similar to the NDW model.  This effect is illustrated in Figure \ref{figL4}.  A comparison between the upper plot without the new rates and the lower plot with the new rates indicates  that the new rates have helped to fill in the discrepancies in abundances below and above the $A=130$ $r$-process peak.  However, this calculation was based upon a slow wind, not a rapid jet model.  Since the wind models are now unviable as a means to produce the heaviest $r$-process nuclei, this does not solve the underproduction problem.

Similarly, \cite{Kratz14} performed  $r$-process calculations in a parameterized (slower transport) NDW scenario based upon the models of \cite{Freiburghaus99}.  Making use of new nuclear  masses and  beta-decay rates  from the finite-range droplet model
FRDM-(2012) \citep{Moller12} it was shown  that the previous discrepancies near $A=120$ are  significantly diminished compared to the same calculation based upon the previous FRDM-(1992) \citep{Moller95} nuclear properties. Hence, one must keep in mind that at least some of the apparent discrepancy may be due to the adopted nuclear input.  Indeed, this is a place where new measurements of nuclear masses near the $r$-process path have made an impact.  In this case, the impact of the new beta-decay rates is to favor models with a more gradual freezeout near the end of the $r$ process.

As another example, calculations of \cite{Nishimura06} could fill the dips in an MHDJ model by using the ETFSI mass model.  However, these models did so at the cost of displacing the 2nd and 3rd peaks and/or  underproducing (or overproducing) abundances over a wide mass region between the second and third peaks.

\subsection{Impact on Models for the $r$ Process in Neutron Star Mergers}
In \cite{Shibagaki16} it was suggested that a solution to the underproduction
of nuclei above and below the $r$-process abundance peaks can be obtained if one considers that both CCSNe and NSMs have contributed to the Solar-System $r$-process abundance distribution.  Indeed, this solution not only resolves the dilemma of underproduction near the peaks, but may help to quantify the relative contributions of CCSNe vs.~NSMs to the Solar-System $r$-process 
abundance distribution.  
   However, this conclusion is very sensitive to the model for fission yields and beta-induced fission as described below.


In \cite{Shibagaki16}  $r$-process simulations were carried out in the NSM model based upon the  merger outflow models of  \cite{Korobkin12,Piran13,Rosswog13}.  These were compared with abundances in the ejecta from the MHD supernova jet model of  \cite{Nishimura12} as well as weak $r$-process yields from the NDW models of \cite{Wanajo13}.
The NSM nucleosynthesis calculations  were evolved using an updated version of the  nuclear network code of  \cite{Terasawa01} with nuclear  masses  from the KTUY model \citep{Koura05} that have been shown  \citep{Nishimura11,Lorusso15} to reasonably well reproduce recently measured beta-decay half-lives of neutron-rich  nuclei and also measured  fission fragment distributions \citep{Ohta07}.

\begin{figure}[htbp]
 \centering
 \includegraphics[scale=0.30,angle=-90]{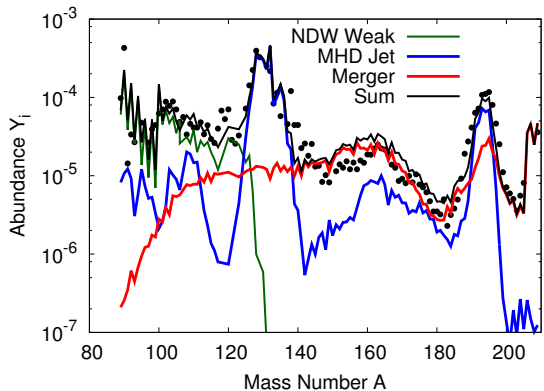}
  \caption{Average final abundance patterns [from \cite{Shibagaki16}] for NSMs (red line), MHDJ (blue line) and the NDW weak $r$-process (green line) from \cite{Wanajo13}.  These are compared with the  Solar-System \citep{Arlandini99} $r$-process abundances (black dots).  The thin black line shows the sum of all contributions.}
 \label{fig5}
\end{figure}

 The red line on Figure \ref{fig5} from \cite{Shibagaki16} shows the result of their  NSM nucleosynthesis simulation   summed over all trajectories of material ejected from the binary NSM simulation.  This is compared with the abundances in the ejecta from the MHD jet   $r$-process (blue line) from  \cite{Nishimura12}, and also  the NDW weak $r$-process abundances (green line) produced in the neutrino driven wind  from the 1.8 M$_\odot$ supernova core calculation of \cite{Wanajo13}.  

The key point of this figure is the possible  role that each process plays in producing the abundance pattern of Solar-System $r$-process abundances [black dots \citep{Arlandini99}].
The total abundance curve from all processes is shown as the black line on Figure \ref{fig5}.  The abundances from  each process were normalized by weighting factors $f_{NSM}$  for NSMs and $f_{Weak}$ for the NDW relative to the MHDJ  yields that were normalized to the $r$-process abundance peaks.
The best fit (black) line in Figure \ref{fig5} is for $f_{NSM} = 0.3$ and $f_{Weak} = 5$.  

These relative contributions  are more or less consistent with Galactic event rates and expected mass ejection from the models.
It was shown in \cite{Shibagaki16}  that weight parameters $f_{NSM}$ and $f_{Weak}$ can be deduced from observed Galactic event rates and expected yields, i.e.~
\begin{equation}
f_{NSM} \approx \frac{R_{NSM} {\rm M_{r, NSM}} }{ \epsilon_{MHDJ} R_{CCSN} {\rm M_{r, MHDJ} }} ~~,
\end{equation}
 and 
 \begin{equation}
f_{Weak} \approx  \frac{R_{CCSN} {\rm M_{r, Weak} }}{\epsilon_{MHDJ} R_{CCSN} {\rm M_{r, MHDJ} }}~~,
\end{equation}
where $\rm M_{r, NSM}$, $\rm M_{r, MHDJ}$, and  $\rm M_{r, Weak}$ are the ejected mass of r-elements from the NSM, MHDJ, and NDW weak-$r$-process models, respectively, while $\rm R_{CCSN}$ and $\rm R_{NSM}$ are the corresponding  Galactic event rates of CCSNe and NMSs.  The quantity $\epsilon_{MHDJ}$ is the fraction of CCSNe that result in magneto-rotationally driven jets. This was estimated  in \cite{Winteler12} to be "perhaps $\sim 1$\%" of the core-collapse supernova rate.  However this is probably uncertain by at least a  factor of two.

The mass of synthesized $r$-process elements from  magneto-rotationally driven jets has been estimated \citep{Winteler12}  to be 6$\times$10$^{-3}M_{\odot}$ while that of a typical  binary NSM is expected to be $(2 \pm 1)\times10^{-2}M_{\odot}$ \citep{Korobkin12}.
If the Galactic NSM rate is $80^{+200}_{-70}$ Myr$^{-1}$ \citep{Kalogera04}, and the Galactic supernova rate is, $(1.9 \pm 1.1) \times 10^4$ Myr$^{-1}$ \citep{Dhiel06}, then it was estimated \citep{Shibagaki16} that  $f_{NSM} \sim 0.6 \pm 0.4 $  and $f_{Weak} \approx  8 \pm 6$ consistent with their  fit parameters.  Although the derivation in \cite{Shibagaki16} is quite uncertain it at  least supports the plausibility of this approach.

Of particular relevance is that  the  one order of magnitude underproduction of nuclides above and below the $A=130$ $r$-process peak from the MHD jet model (shown by the blue line) is nearly accounted for by contributions from  the NDW and NSM models.

The final $r$-process isotopic  abundances from the NSM model of \cite{Shibagaki16} exhibited a  very flat pattern due to several episodes of fission cycling under extremely neutron-rich conditions. 
Thus, NSMs may resolve most of the underproduction problems of  the MHDJ model predictions for the elements just below and above the abundance peaks and be largely responsible for the
rare-earth abundances in the range $A = 140-180$.  
The remaining underproduction below the $A=130$ peak could then be due to the NDW weak $r$-process as illustrated on Figure \ref{fig5}.

The conclusions of \cite{Shibagaki16}, however, were critically dependent upon the fission barriers and fragment distributions adopted in that study.
An  important  difference between  the work of \cite{Shibagaki16} and that of other NSM studies \citep{Goriely11, Korobkin12,  Piran13, Rosswog13,Rosswog14,Goriely13,Wanajo14, Nishimura16}  is the termination of the $r$-process path.  

\begin{figure}[htbp]
 \centering
 \includegraphics[scale=0.55]{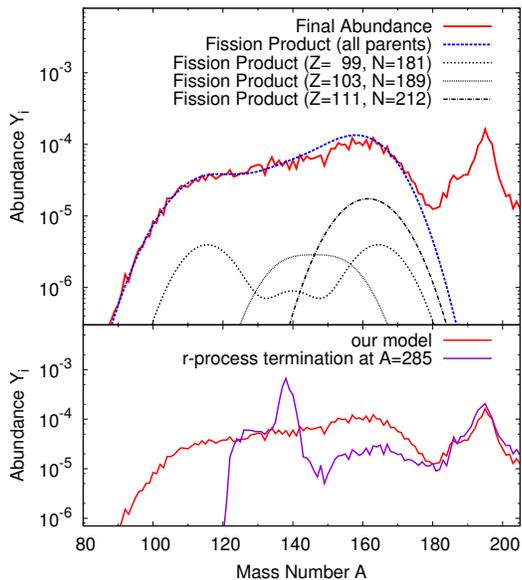}
  \caption{Illustration of the impact of fission yields and fission recycling  on the final $r$-process abundances from \cite{Shibagaki16}. Upper panel shows the relative contributions for 3 representative nuclei compared with the final abundance distribution.    The lower panel shows the same final $r$-process yields  compared with the distribution that would result if the termination of the $r$-process path were to occur at $A=285$.    }
 \label{fig6}
\end{figure}

The $r$-process path in the  NSM calculations of \cite{Shibagaki16} proceeds rather below the fissile region until nuclei with $A \sim 320$, whereas the $r$-process path based upon microscopic calculations of fission barriers [such as  \cite{Goriely13}]  terminates at $A \approx 278$ [or for a maximum $\langle Z \rangle$ for \citep{Korobkin12}]. Also, in \cite{Shibagaki16} only $\sim10$\% of the final yield came from the termination of the $r$-process path at $N = 212$ and $Z = 111$,  while almost 90\% of the $A \approx160$ bump shown in Figure \ref{fig5} was from the fission of  more than 200  different parent nuclei mostly via beta-delayed fission.  This is illustrated in the upper panel of Figure \ref{fig6} from \cite{Shibagaki16}.

This is in contrast to the yields of \cite{Goriely13} that are almost entirely due to a few  $A \approx 278$ fissioning nuclei with a characteristic four hump FFD.  It is important to keep in mind that many  microscopic calculations of fission barriers are consistent with an early termination near $A \approx 280$.  It is only the phenomenological fission probabilities   \citep{Koura04,Chiba08} that allow the $r$-process path to continue to $A >300$. The difference in $r$-process abundances from the two possibilities is significant.

This point is illustrated  in the lower panel of Figure \ref{fig6}.  This figure compares with a calculation in which it was  assumed  that the $r$-process path was terminated by symmetric fission of  nuclei with $A=285$ [similar to that of \cite{Goriely13}].  In this case a solar-like distribution is obtained similar to that of  \cite{Goriely11,Korobkin12,Goriely13}.  This highlights the importance of eventual detailed measurements of  fission barriers and fragment distributions for nuclei near the termination of the  $r$-process path.
   
\section{Impact on Models for Galactic Chemical Evolution}
\label{GCE}

In this review we  suggest that the measured and best estimates of nuclear input masses and beta-decay rates near the $r$-process path support an $r$-process model in which the freezeout of abundances is slower than that of the MHD jet models.  This may support a large contribution from NSMs.  Although NSMs appear to make a good $r$-process site, chemical evolution studies \citep{Mathews90,Argast00,Argast02,Argast04} have shown that very  short merger times of NSMs are needed to reproduce  abundances seen in  $r$-enhanced  extremely metal-poor stars in the Galactic halo. 

There are two possible scenarios for a large contribution from NSMs.  One is that the first stars had a large contribution from MHDJ ejecta that was later supplanted by the contribution from NSMs.  This possibility was considered in \cite{Shibagaki16} where it was shown that the universality in {\it elemental} abundances for the low metallicity first stars is 
consistent with the MHD jet yields.  This is  because the isotopic sum to produced elemental abundances tends to fill in the missing isotopes above and below abundance peaks.  However, some slight deviations from universality  may appear in this scenario \citep{Shibagaki16}.  It would be valuable to look for evidence for such deviations in the most metal-poor $r$-enhanced stars.

Another possibility is that the time scale of metal enrichment in the most metal-poor stars does not follow a simple age-metallicity relation.  That is, some of the enrichment of $r$ elements may occur in dwarf galaxies that later merge with the Galactic halo.

Recently,  \cite{Hirai15} considered the  chemical evolution in dwarf spheroidal galaxies (dSphs).  Such dSphs  are the  building blocks of the Galactic halo and have a much lower star formation efficiency than that of the Milky way halo. That paper showed  that  when the effect of metal mixing was taken into account,  the enrichment of $r$-process elements in dSphs by NSMs could reproduce the observed [Eu/Fe] vs.~metallicity distribution with a merger delay time time of as much as 300 Myr. This is because metallicity is not really correlated with the time $\sim 300$ Myr from the start of the simulation to the low star formation efficiency in dSphs. They  also confirmed that this model is consistent with observed properties of dSphs such as the radial profiles and metallicity distribution. A merger time  of $\sim 300$ Myr and a Galactic NSM rate of $\sim 10^{-4}$ yr$^{-1}$ could reproduce the abundances of metal poor $r$-process enhanced stars and is consistent with the values suggested by population synthesis and other nucleosynthesis studies.

\section{Summary of the most needed nuclear measurements}
\label{needed}

\begin{figure}[htbp]
 \centering
 \includegraphics[scale=0.3]{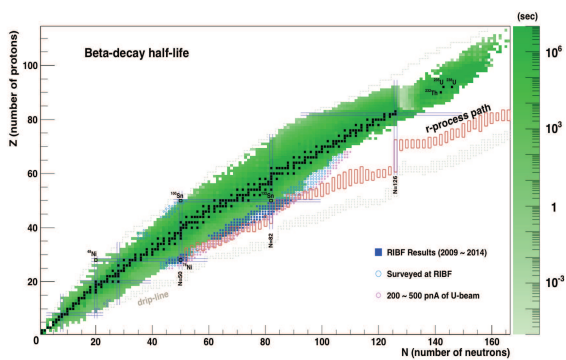}
  \caption{Illustration of the isotopes along the r-process path indicating those beta half lives that have been measured up to 2014 at RIKEN and those that may be accessible via a high intensity uranium beam in the future.   }
 \label{nishimurafig}
\end{figure}

From the point of view of this review there are two main thrusts where new data would be most helpful.  For understanding the MHDJ model, it is imperative to know the beta-decay rates above and below the r-process peaks at A=130 and 195.  Figure \ref{nishimurafig} (S. Nishimura, Priv. Comm.) summarizes the current situation of measurements at RIKEN until 2014 and what is accessible with a higher intensity uranium beam.  This figure shows that a number of isotopes along the r-process path near the $N=82$ closed neutron shell now have measured beta-decay lifetimes. However, a number of isotopes above the shell still need investigation.  Moreover, for all of these nuclei precise nuclear masses must be determined, perhaps via beta-decay endpoint measurements.  It is important to know the neutron separation energies to quantify the degree of shell quenching in this region.   The reader is referred to  \cite{Brett12} for a list of the most important separation energies to measure near the $N=82$ closed neutron shell and also \cite{Mumpower16} for an exhaustive list of the most important isotopes and measurements  in the context of various $r$-process paradigms.  Also, nuclear spectroscopy needs to be completed for these isotopes to determine the nuclear partition functions.

In particular, beta decay lifetimes along the r-process path in the rare earth peak region and near the A=195 peak  are crucial measurements.   Even the simple question as to whether  the r-process occurs in a cold or hot, or fission recycling environment could be answered if beta decay rates and masses are known near the rare-earth peak at $A \sim 160$ \citep{Mumpower16}. In particular, nuclear properties of a few isotopes in the range of $Z \approx 53-60, N \approx 100-115$ can have a dramatic effect on the final freezeout abundances for the rare-earth peak \citep{Mumpower16}.  
Eventually, beta-decay lifetimes and nuclear masses near the $A =195, N=126$ neutron closed shell are also desired as a means, in particular, to test the viability of the MHDJ model.

As also outlined in \cite{Mumpower16} neutron capture rates along the $r$-process path are crucial for some isotopes near the $A=130$ and $A=195$ peaks and also near the rare-earth peak at $A\sim 160$.  As noted above, what is needed is an ambitions project for direct measurements using inverse kinematics and a storage ring, or indirect $(d,p)$ and virtual $(\gamma,n)$ measurements via Coulomb excitation of unstable beams.

Regarding neutron star mergers, it is critical to understand fission barriers, beta-induced fission rates and fission fragment mass distributions in the vicinity of the heaviest $A \sim 280-300$ nuclei near the termination of the r-process path.  If fission barriers are low so that fission occurs via one or two  isotopes near $A \sim 285$, and if this fission produces a bimodal distribution
as microscopic calculations suggest \citep{Goriely13}, then NSMs are a very viable candidate to produce the entire r-process abundances.  On the other hand, if the r-process proceeds all the way to nuclei with $A \sim 300$ as in the study of \cite{Shibagaki16}, then NSMs may only contribute to the rare-earth peak plus help to fill the gaps above and below the r-process peaks.  
The best resolution of this question would be to directly measure fission barriers and beta-delayed fission in this region.  This is a difficult measurement to make, but perhaps the formation of these nuclei in via a radioactive-ion reaction followed by beta decay into the region of fissile nuclei could be used to reveal the fission barriers and fission fragment mass distributions.

\section{Conclusions}
\label{concl}
In this review we have considered the various models for $r$-process nucleosynthesis and how new measurements of $\beta$-decay rates and nuclear masses near the $r$-process path have
impacted these models.
A main impact of the new measurements concerns the tendency of models with a rapid freezeout timescale to underproduce isotopes below and above the main $r$-process abundance peaks.

Although the new mass data suggest some shell quenching around the neutron closed shells far from stability, the indication from the $\beta$-decay rates suggests that this quenching is not enough to prevent the underproduction in models with a rapid freezeout like MHD jets.  Phenomenological models with a more gentle  freezeout (as in the NDW) seem to best reproduce the $r$-process abundances.  However, since NDW models are out of favor on theoretical grounds, the need for NSM contributions to the $r$ process is apparent.  Although material is tidally ejected at high velocity in NSMs, the freezeout can occur on a more gradual timescale in the frame of the ejected material due to the very high neutron density.  

The question remains, however, as to whether the fission recycling environment of NSMs can reproduce all of a part of the $r$-process abundance distribution.  The answer to that question will require the continual accumulation of masses, beta-decay rates, and in particular, fission barriers, and fission mass distributions for the heaviest neutron-rich nuclei near the termination of the $r$-process path.

\begin{acknowledgments}
This study was supported in part by Grants-in-Aid
for Scientific Research of JSPS (19340074, 20244035, and 22540290),
JSPS Fellows (21.6817), Scientific Research on Innovative Area of MEXT (20105004),
and U.S. Department of Energy under Nuclear Theory Grant DE-FG02-95- ER40934.
This work also benefited from support by the National Science Foundation under Grant No. PHY-1430152 (JINA Center for the Evolution of the Elements).
\end{acknowledgments}


\end{document}